\documentclass[pra,reprint,aps,twocolumn,showpacs,superscriptaddress,floatfix,amsmath,amssymb,nofootinbib]{revtex4-2}
\usepackage{amsmath,amssymb,graphicx,braket,bbold}
\usepackage{physics}
\usepackage{hyperref}
\usepackage{xcolor}
\usepackage{graphicx}
\usepackage{enumitem}
\usepackage{ulem}
\newcommand{\beq}{\begin{equation}}
\newcommand{\eeq}{\end{equation}}
\newcommand{\beqa}{\begin{eqnarray}}
\newcommand{\eeqa}{\end{eqnarray}}

\newcommand{\appropto}{\mathrel{\vcenter{
  \offinterlineskip\halign{\hfil$##$\cr
    \propto\cr\noalign{\kern2pt}\sim\cr\noalign{\kern-2pt}}}}}

\hypersetup{colorlinks,citecolor=cyan,filecolor=blue,linkcolor=blue,urlcolor=blue}

\begin{document}

\title{Enhancing polarization transfer from nitrogen-vacancy centers in diamond to external nuclear spins via dangling bond mediators}
\author{H. Espin\'os}
\email{hespinos@fis.uc3m.es}
\affiliation{Departamento de F\'isica, Universidad Carlos III de Madrid, Avda. de la Universidad 30, 28911 Legan\'es, Spain}
\author{C. Munuera-Javaloy}
\affiliation{Department of Physical Chemistry, University of the Basque Country UPV/EHU, Apartado 644, 48080 Bilbao, Spain}
\affiliation{EHU Quantum Center, University of the Basque Country UPV/EHU, Leioa, Spain}
\author{I. Panadero}
\affiliation{Departamento de F\'isica, Universidad Carlos III de Madrid, Avda. de la Universidad 30, 28911 Legan\'es, Spain}
\affiliation{Arquimea Research Center, Camino las Mantecas s/n, 38320 Santa Cruz de Tenerife, Spain}
\author{P. Acedo}
\affiliation{Department of Electronic Technology, University Carlos III de Madrid, Avda. de la Universidad 30, 28911 Legan\'es, Spain}
\author{R. Puebla}
\affiliation{Departamento de F\'isica, Universidad Carlos III de Madrid, Avda. de la Universidad 30, 28911 Legan\'es, Spain}
\author{J. Casanova}
\affiliation{Department of Physical Chemistry, University of the Basque Country UPV/EHU, Apartado 644, 48080 Bilbao, Spain}
\affiliation{EHU Quantum Center, University of the Basque Country UPV/EHU, Leioa, Spain}
\affiliation{IKERBASQUE,  Basque  Foundation  for  Science, Plaza Euskadi 5, 48009 Bilbao,  Spain}
\author{E. Torrontegui}
\affiliation{Departamento de F\'isica, Universidad Carlos III de Madrid, Avda. de la Universidad 30, 28911 Legan\'es, Spain}
%\email{eriktorrontegui@gmail.com}

\begin{abstract}

The use of nitrogen-vacancy (NV) centers in diamond as a non-invasive platform for hyperpolarizing nuclear spins in molecular samples is a promising area of research with the potential to enhance the sensitivity of nuclear magnetic resonance (NMR) experiments. Transferring NV polarization out of the diamond structure has been achieved on nanoscale targets using dynamical nuclear polarization methods, but extending this polarization transfer to relevant NMR volumes poses significant challenges. One major technical hurdle is the presence of paramagnetic defects in the diamond surface which interfere with polarization outflow. However, these defects can be harnessed as intermediaries for the interaction between NVs and nuclear spins. We present a method that benefits from existing microwave sequences, namely the PulsePol, to transfer polarization efficiently and robustly using dangling bonds or other localized electronic spins, with the potential to increase polarization rates under realistic conditions.
\end{abstract}

\maketitle
\section{Introduction}
Nuclear magnetic resonance (NMR) is a versatile and powerful technique applied across diverse fields, ranging from material physics~\cite{Lilly2017} to life-sciences~\cite{Wu2016}, owing to its analytical properties, as well as to its non-invasive character.
In particular, NMR-based detection techniques rely on the nuclear spin polarization of the target sample, which is minute at room temperature --namely, on the order of $10^{-5}$ at externally applied magnetic fields of 2 T~\cite{Schwartz2018}--, thus leading to low sensitivity issues. For this reason, NMR spectroscopy highly benefits from dynamical nuclear polarization (DNP)~\cite{Kuhn2013} methods that aim to transfer polarization from electron spins to nuclei,   
leading to an increase of the NMR signal of several orders of magnitude~\cite{Ardenkjaer2003}. Typically, DNP methods bridge the energy gap between electrons and nuclei via microwave (MW) irradiation, thus enabling the transfer of thermally polarized electrons to the nuclear environment.

In this scenario, nitrogen vacancy (NV) centers in diamond~\cite{Doherty2013} are a promising polarization device, as their electron-spin degrees of freedom can be initialized/polarized to a degree higher than $90\%$ at room temperature via green laser irradiation~\cite{Jelezko2006,Waldherr2011}. Consequently, different DNP methods have been designed and experimentally realized with the objective of transferring the optically induced NV polarisation to the nearby $^{13}$C nuclear spins in the diamond lattice. This has been achieved both in bulk diamond~\cite{Schwartz2018,London2013,King2015,Pagliero2018}, as well as in nanodiamond particles~\cite{Chen2015,Ajoy2018,Henshaw2019}. Transferring NV polarization out of the diamond  has been achieved on nanoscale targets~\cite{Shagieva2018,Broadway2018,Rizzato2022}, but extending DNP methods to sample volumes used in standard NMR presents serious difficulties. Among these challenges, system inhomogeneities, such as energy shifts and microwave driving deviations, threaten to the efficacy of DNP methods. Furthermore, a significant impediment lies in the physical separation between the sample and the NV spins, which are confined within diamond crystals, meaning that even the NV spins closest to the diamond surface are situated several nanometers away from the sample. This separation is notably greater than the distances on the order of 1 nm or less typically encountered in conventional DNP, where polarizing agents and nuclear spins are intimately mixed on the molecular scale~\cite{Lilly2017,Plainchont2018}. Additionally, paramagnetic defects on the surface can interfere with the polarization outflow~\cite{Rosskopf2014,Romach2015,Sangtawesin2019,Hall2020}.

These paramagnetic defects can be treated as mediators of the interactions between NVs and other nuclear spins. A recent experiment demonstrated polarization transfer to $^{13}$C nuclei assisted by P$_1$ centers using MW-free DNP adiabatic processes~\cite{Henshaw2019}. Additionally, some of the surface defects, located within a few layers of the diamond facet, give rise to dangling bonds with spin 1/2 and the g-factor of a free electron. 
These surface spins have been shown to be chemically stable in ambient conditions over many days~\cite{Lukin2014}. Different approaches have used this single electron spin as mediator
for sensing purposes ~\cite{Lukin2014,Schaffry2011,Bleszynski2022}.
Regarding spin polarization, the development of MW sequences that transfer polarization from NV to nuclei in a fast and robust manner using dangling bonds (or other localized electron-like mediators via functionalized surfaces) has the potential to accelerate the polarization enhancement of bulk samples.

Commonly employed protocols encounter substantial challenges when attempting to transfer polarization to spins with high Larmor frequencies (typically higher than a hundred of MHz). The Larmor frequency is the product of the applied magnetic field and the gyromagnetic ratio of the particle in such a magnetic field. Therefore, particles subjected to large magnetic fields, specially those with significant gyromagnetic ratios, such as electrons, are affected by limitations that depend on the specific polarization protocol. For example, in Nuclear spin Orientation Via Electron spin Locking (NOVEL)\cite{Henstra2008}, successful polarization transfer requires the microwave spin-lock amplitude to match the Larmor frequency. This is not always feasible when the Larmor frequency of the target spin exceeds the amplitude achievable with a driving source without causing sample damage. In the PulsePol sequence\cite{Schwartz2018}, the free evolution time between consecutive pulses is inversely proportional to the target's Larmor frequency. Yet, due to the finite pulse duration, large Larmor frequencies can lead to pulse overlap. These constraints often confine the application of polarization sequences to nuclei with small gyromagnetic ratios experiencing low magnetic fields.

In this paper, we show that, by applying a double-channel PulsePol sequence to simultaneously address both an NV spin and a surface electron spin, the aforementioned restrictions on polarization transfer to spins with high Larmor frequencies are lifted. The resulting polarization transfer protocol is resistant to a wide range of control errors due to the high degree of robustness of the PulsePol sequence. Then, by adjusting the free time between pulses to the Larmor frequency of an external nucleus, we will 
demonstrate that polarization is further transferred from the NV to external nuclei, using the surface electron spin as a mediator (see Fig.~\ref{fig:sequence}(a) for a schematic representation of the system that will be analyzed).

The paper is organized into five more sections. In Section \ref{section:I}, we present the theoretical framework for describing the polarization transfer using the double-channel PulsePol sequence. In Section \ref{section:II}, we extend this framework to describe the transfer of polarization to a nucleus, using a surface electron spin as an intermediary. The simulated results are shown in Sec. \ref{section:results}. Section \ref{section:III} contains simulations of the polarization transfer to different target systems using a master equation formalism, and we analyze the robustness properties of the sequence. Additionally, we introduce an approximate trend for the polarization buildup as a function of the couplings based on \textit{cooling rates}. In Section \ref{section:IV}, we apply these cooling rates to study the polarization transfer in a continuum description using a convection-diffusion differential equation. Finally, the conclusions of our work are summarized in Section \ref{section:V}.

{\section{Double-channel PulsePol sequence}\label{section:I}
The optically induced NV spin polarization, surpassing a degree of 90\%, provides a sustainable and continuously renewable source for polarizing nearby spin ensembles. However, given that the distance between NV centers and nuclei external to the diamond is of several nanometers, the dipole-dipole coupling that governs the polarization transfer is weak. Bringing NV centers near the diamond surface presents a significant challenge, as fast-fluctuating surface defects negatively impact the NV coherence time $T_2$, which is already a limiting factor for polarization transfer~\cite{Tetienne2021}. 

A recent study showed the existence of isolated electronic spins or dangling bonds on the surface of high purity diamond, with the potential of being coherently manipulated~\cite{Lukin2014,Rezay2022}. These electronic spins are localized with nanometer-size uncertainty. Multitude of NV measurements have determined that these spins baths have a g-factor of 2, and areal densities of $\sigma\sim 0.01-0.5$ nm$^{-2}$~\cite{Rosskopf2014,Lukin2014,Myers2014}. Despite being close to the surface, they are stable in air over time scales of many days. Another study~\cite{Chou2021} demonstrated that the Fermi-contact term in the hyperfine coupling is not negligible between the surface spins and the surrounding nuclear spins, and thus there is a considerable interaction between them. Sushkov et al.\cite{Lukin2014} used these electronic spins as magnetic resonance ``reporters" for sensing and imaging of individual proton spins lying also at the surface. 

Due to the substantially larger gyromagnetic factor of electrons compared to nuclei commonly utilized in NMR experiments, the resulting dipole-dipole coupling between the NV center and an electron is significantly stronger than the NV-nuclei interaction. In the following, we introduce a method that takes advantage of the strong NV-electron coupling, together with the close proximity of this electron to external nuclei, to bring polarization out of the diamond lattice containing the NV center, using the electron as mediator. To overcome the challenge of transferring polarization to a spin with a high Larmor frequency, we employ a double-channel sequence that concurrently acts on both the NV center and the electron. This innovative technique allows for efficient polarization transfer without necessitating resonance conditions matching the Larmor frequency of the surface electron.

%%%%%%%%%%%%%%%%%%%%%%%%%%%%%%%%%%%%%%%%%%
\begin{figure*}[t!]
    \centering
    \includegraphics[width=1\linewidth]{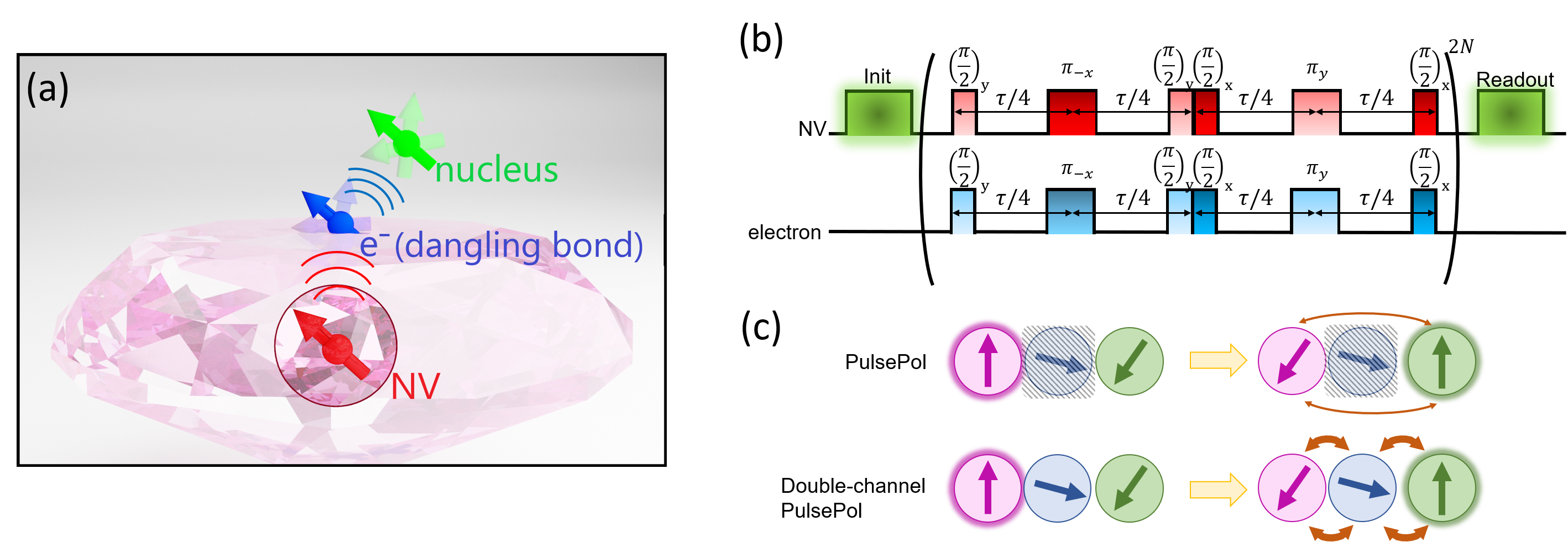}
  \caption{Schematics of the proposed protocol. (a) The nitrogen-vacancy (NV) spin state is transferred through an intermediate electron at the diamond surface. (b) Schematics of the microwave pulse sequence on the NV and the electron spin for pulsed polarization transfer. (c) Comparison between the direct PulsePol scheme and the double-channel PulsePol. In the latter, the electron interacts coherently with the NV and the nearby nuclei to mediate the polarization process. The introduction of the electron as mediator allows for stronger interactions and thus faster protocols. }
  \label{fig:sequence}
\end{figure*}
%%%%%%%%%%%%%%%%%%%%%%%%%%%%%%%%%%%%%%%%%%

The Hamiltonian describing the NV, the surface spin, their interaction and a microwave driving is
\begin{equation}\label{eq:H}
    H = D S_\text{z}^2-\gamma_\text{e} B  S_\text{z}-\gamma_\text{e} B E_\text{z} + \mathbf{S}\cdot \mathcal{A}\cdot\mathbf{E} + H_\text{D},
\end{equation}
where $D=2\pi\times2.87$ GHz is the zero-field splitting of the NV, $\gamma_\text{e}$ is the gyromagnetic ratio of the electron, $B$ is the magnetic field (aligned with the NV quantization axis), $\mathbf{S}$ $(\mathbf{E})$ is the spin--$1\left(\frac{1}{2}\right)$ operator of the NV (electron), and $\mathcal{A}$ is the electron-electron dipole coupling tensor describing the interaction between the NV and the electron resulting from the magnetic dipole-dipole coupling. $H_\text{D}$ describes the interaction with two external microwave sources affecting the NV and the electron, 
\begin{eqnarray}\label{eq2}
    H _\text{D}&=& \Omega_1(t)(\sqrt{2}S_\text{x}+2E_\text{x})\cos (\omega_1 t + \varphi_1) \\ \nonumber
    &+&\Omega_2(t)(\sqrt{2}S_\text{x}+2E_\text{x}) \cos(\omega_2 t+\varphi_2),
\end{eqnarray}
where $\Omega_i$, $\omega_i$ and $\phi_i$ denote the Rabi frequency, the frequency, and the phase of the $i$-th driving field, respectively.
These drivings allow us to simultaneously control the NV and the electron by sending pulses with $\omega_1$ in resonance with one of the NV's transitions ($\omega_1=D\pm \abs{\gamma_\text{e}}B+\Delta_{\rm NV}$, with $\Delta_{\rm NV}$ a small detuning) and $\omega_2$ in resonance with the electron's transition ($\omega_2=\abs{\gamma_\text{e}}B+\Delta_{\rm e}$, with $\Delta_{\rm e}$ another small detuning). As long as $\omega_1-\omega_2 \gg \Omega_1,\Omega_2$, the first term in Eq.\eqref{eq2} will only produce transitions in the NV, while the second term will only produce them in the electron. This condition is automatically satisfied if the NV $\ket{0}\leftrightarrow\ket{1}$ transition is selected, since then $\omega_1-\omega_2 \approx D = 2\pi\cross 2.87$ GHz, much greater than the Rabi frequencies (in the order of tens of MHz). Thus, going to a rotating frame with respect to $H_0= D S_\text{z}^2-(\gamma_\text{e} B-\Delta_\text{NV}) S_\text{z}+\omega_2 E_\text{z}$, we get, after applying the rotating wave approximation to eliminate fast rotating terms (see Supplementary Note 1),
\begin{eqnarray}\label{eq:H_RWA}
    H &\simeq &S_\text{z}\mathcal{A}_\text{zz}E_\text{z} + \Delta_{\rm NV}S_\text{z}+\Delta_{\rm e}E_\text{z} \nonumber\\
    &+& \frac{\Omega_1(t)}{2}\left(\sigma_\text{x}^\text{NV}\cos\varphi_1-\sigma_\text{y}^\text{NV}\sin\varphi_1\right)\nonumber\\
    &+&\Omega_2(t)\left(E_\text{x}\cos\varphi_2+E_\text{y}\sin\varphi_2\right),
\end{eqnarray}
where we have introduced the operators $\sigma_\text{x}^\text{NV} = \ketbra{1}{0}+\ketbra{0}{1}$ and $\sigma_\text{y}^\text{NV} = -i(\ketbra{1}{0}-\ketbra{0}{1})$.
This approximation is valid when $\omega_1 - \omega_2$ is also much greater than any term coming from the electron-electron dipole coupling tensor. The coupling term is given by
\begin{equation}\label{eq:Azz}
    \mathcal{A}_\text{zz}=\frac{\hbar\mu_0\gamma_\text{e}^2}{4\pi\abs{\mathbf{r}}^3}(1-3\cos^2\theta),
\end{equation}
where $\mathbf{r}$ is the position vector of the electron with the NV at the origin and $\theta$ is the angle that this vector forms with the magnetic field. For simplicity in the derivations that follow, we assume that the drivings are perfectly tuned ($\Delta_\text{\rm NV}=\Delta_\text{\rm e}=0$), and the $\ket{-1}$ level is ignored, since it is not affected by the dynamics. Therefore, we reduce the NV Hilbert space to only 2 levels introducing $\sigma_{z}^\text{NV}=\ketbra{1}{1}-\ketbra{0}{0} =2S_\text{z}- \mathbb{1}$. Thus, the Hamiltonian \eqref{eq:H_RWA} can then be rewritten as
\begin{eqnarray}\label{eq:H_RWA2}
    H &\simeq &\frac{1}{2}\mathcal{A}_\text{zz}(\sigma_\text{z}^\text{NV}E_\text{z}+E_\text{z}) \nonumber\\
    &+& \frac{\Omega_1(t)}{2}\left(\sigma_\text{x}^\text{NV}\cos\varphi_1-\sigma_\text{y}^\text{NV}\sin\varphi_1\right)\nonumber\\
    &+&\Omega_2(t)\left(E_\text{x}\cos\varphi_2+E_\text{y}\sin\varphi_2\right).
\end{eqnarray}

The polarization protocol is designed to transform the ZZ-interaction from Eq.~\eqref{eq:H_RWA} into a flip-flop interaction, which is the sum of an interaction of the type $\sigma_\text{x}^\text{NV}E_\text{x}$ and another one of the type $\sigma_\text{y}E_\text{y}$. The basic sequence block is applied simultaneously to the NV and the electron, as shown in Fig.~\ref{fig:sequence}(b), and is given by
\begin{equation}\label{eq:sequence}
    \left[\left(\frac{\pi}{2}\right)_\text{Y}\rule{10pt}{0.1pt}\left(\pi\right)_{-\text{X}}\rule{10pt}{0.1pt}\left(\frac{\pi}{2}\right)_\text{Y}\left(\frac{\pi}{2}\right)_\text{X}\rule{10pt}{0.1pt}\left(\pi\right)_\text{Y}\rule{10pt}{0.1pt}\left(\frac{\pi}{2}\right)_\text{X}\right]^{2},
\end{equation}
where $\left(\phi\right)_{\pm \text{X,Y}}$ denote $\phi-$pulses around the $x/y$ axes with duration $t'_{1,2}=\phi/\Omega_{1,2}$. Unlike the PulsePol applied to just one of the participating elements, the time between pulses, during which the NV and electron undergo free evolution, is completely arbitrary ($\tau/4$ in Fig.~\ref{fig:sequence}(b)). The sequence can be divided into two blocks. In each block, $\pi/2$-pulses are applied simultaneously to both the NV and the electron to transform the ZZ-interaction into either an XX- or a YY-interaction. A $\pi$-pulse is introduced in the middle of each block to eliminate the $\frac{\mathcal{A}\text{zz}}{2}E_\text{z}$ term in Hamiltonian~\eqref{eq:H_RWA2}, while at the same time canceling other detuning errors accumulated during the free evolution. Therefore, the evolution under Hamiltonian~\eqref{eq:H_RWA2} can be rewritten as (see Supplementary Note 2)
\begin{equation}\label{eq:evolution_operator}
    U_{seq} = \exp\left(-i\frac{\tau}{2}\frac{\mathcal{A}_\text{zz}}{2}\sigma_\text{x}^\text{NV}E_\text{x}\right)\exp\left(-i\frac{\tau}{2}\frac{\mathcal{A}_\text{zz}}{2}\sigma_\text{y}^\text{NV}E_\text{y}\right),
\end{equation}
where $\tau$ is the duration of one sequence. When only one NV and one electron are involved, or for sufficiently small $\tau$ ($\tau\ll 1/\mathcal{A}_\text{zz}$), the evolution becomes
\begin{equation}\label{eq:evolution_operator2}
    U_{seq}= \exp\left[-i\frac{\tau}{2}\frac{\mathcal{A}_\text{zz}}{2}\left(\sigma_\text{x}^\text{NV}E_\text{x}+\sigma_\text{y}^\text{NV}E_\text{y}\right)\right].
\end{equation}
Notice that, in the single PulsePol, the effective coupling is reduced to at most $\sim 72\%$ of the coupling strength. However, in the double-channel PulsePol sequence, the full coupling strength takes part in the polarization rate. 
In a system in which the NV is initially polarized in the $\ket{0}$ state and the electron is in a thermal state, the polarization $P$ after a sequence becomes
\begin{equation}\label{eq:pol1}
P(\tau)=\\Tr\left[U_\text{seq}\rho(0)U_\text{seq}^\dagger 2E_\text{z}\right]=-\sin^2\left(\frac{\mathcal{A}_\text{zz}\tau}{4}\right),
\end{equation}
where, if we denote $p_\uparrow$ and $p_\downarrow$ as the probabilities for the electron spin to be in the $\ket{\uparrow}_\text{e}$ and $\ket{\downarrow}_\text{e}$ states (which are eigenstates of the $E_\text{z}$ operator), respectively, the polarization $P$ is defined as $P=p_\uparrow-p_\downarrow$. The minus sign in Eq.~\eqref{eq:pol1} indicates that the polarization is acquired in the $\ket{\downarrow}_\text{e}$ state. In practice, due to the pulsed nature of the sequence, there is an additional rate reduction of the polarization buildup that arises due to the finite strength of the pulses. During $\pi/2$-pulses, the NV and the electron are not optimally coupled. The effective coupling rate is the time-averaged instantaneous coupling strength over the whole sequence. In particular, if the NV and the electron spin rotate synchronously, i.e., $\Omega_1=\Omega_2\equiv\Omega$, there is a reduction of the effective interaction to half its maximum value during $\pi/2$-pulses.  As a result, the polarization rate is attenuated to (see Supplementary Note 3)
\begin{equation}\label{eq:reduced_coupling}
    \mathcal{A}_\text{zz}\mapsto \mathcal{A}_\text{zz}\left(1-\frac{\pi}{\Omega\tau}\right).
\end{equation}

\section{Polarization of an external nucleus using an intermediate electron}\label{section:II}
The previous polarization transfer mechanism is suitable for any target particle, regardless of its Larmor frequency, and maintains the robustness of PulsePol without requiring a resonance condition on the interpulse spacing. This feature makes it advantageous for transferring the spin state from the NV center to an external nucleus via the mediator electron. The proposed protocol is shown in Fig.~\ref{fig:sequence}. The electron's close proximity to the diamond surface leads to a larger coupling with external nuclei in comparison to the coupling between an NV center within the diamond lattice (shallow NVs are positioned some nanometers beneath the surface) and external nuclear spins. Integrating the flexibility of pulse spacing from the previous protocol with the PulsePol resonance condition enables the concurrent transfer of spin states from the NV center to the electron, and from the electron to the external nucleus. The larger coupling between these elements yields a considerably elevated polarization transfer rate, surpassing the efficiency of direct transfer from the NV center to the external nucleus. A representation of the considered system is shown in Fig.~\ref{fig:sequence}(a). The Hamiltonian describing the whole system is similar to Eq.~(\ref{eq:H}) but includes the Zeeman term of the nucleus and the dipole coupling between the electron and the nucleus,
\begin{eqnarray}
    \label{eq:H2}
    H_T&=& H+\gamma_\text{n}BI_\text{z}+\mathbf{E}\cdot\mathcal{B}\cdot\mathbf{I}\nonumber\\
    &=& D S_\text{z}^2-\gamma_\text{e} B  S_\text{z}-\gamma_\text{e} B E_\text{z} + \gamma_\text{n} B I_\text{z} +\\
    &+& \mathbf{S}\cdot\mathcal{A}\cdot\mathbf{E} + \mathbf{E}\cdot\mathcal{B}\cdot\mathbf{I} + H_\text{D}.\nonumber
\end{eqnarray}
Here, $\mathbf{I}$ is the spin--$\frac{1}{2}$ operator of the external nucleus, $\gamma_\text{n}$ is the gyromagnetic factor of the nucleus, and $\mathcal{B}$ is the coupling tensor  describing the interaction between the electron and the nucleus.  The interaction between the NV center and the nucleus has been neglected in favor of the much larger coupling between the electron and nucleus, assuming that they are in much closer proximity. The driving $H_\text{D}$ is the one given by Eq.~(\ref{eq2}), i.e., there are no additional pulses applied to the nucleus. According to the PulsePol resonance condition, by choosing a pulse spacing such that
\begin{equation}\label{eq:tau}
\tau=\frac{n\pi}{\gamma_\text{n} B},
\end{equation}
with $n$ odd, it is possible to generate effective flip-flop dynamics between the electron and the nucleus. In contrast to the electron spin, the gyromagnetic ratio of nuclei is comparatively smaller. Consequently, the condition described by Eq.~\eqref{eq:tau} is generally not a constraining factor for the sequence's feasibility, even at moderate magnetic fields. It is important to acknowledge, however, that practical considerations must be taken into account. Compensations are necessary due to the finite duration of pulses, leading to a reduction in the free evolution time $t_{\text{free}}$ that ensures that the total block maintains the same overall duration $\tau$. Assuming, for simplicity, that both driving pulses have the same Rabi frequency $\Omega$,
\begin{equation}
    t_{\text{free}} = \tau - 2\frac{\pi}{\Omega} - 4\frac{\pi}{2\Omega} = \tau - 4\frac{\pi}{\Omega}.
\end{equation}
In a frame rotating with respect to $H_0= \omega_1 S_\text{z}+\omega_2 E_\text{z}+\gamma_n B I_z$, discarding all fast-rotating terms, the effective Hamiltonian becomes under the previous condition \cite{Schwartz2018}
\begin{equation}\label{eq:Heff}
    H_{\text{eff}}= \frac{\mathcal{A}_\text{zz}}{4}\left(\sigma_\text{x}^\text{NV}E_\text{x}+\sigma_\text{y}^\text{NV}E_\text{y}\right)+\frac{\alpha\mathcal{B}_{\perp}}{2}\left(E_\text{x}I_\text{x}+E_\text{y}I_\text{y}\right),
\end{equation}
where $\alpha<1$ is a constant determined by the filter function generated by the PulsePol sequence and depends on the harmonic $n$ \cite{Schwartz2018}, and the coupling term $\mathcal{B}_{\perp}$ is given by
\begin{equation}\label{eq:Bzx}
    \mathcal{B}_{\perp} = \frac{3\hbar\mu_0\gamma_\text{e}\gamma_\text{n}}{4\pi\abs{\mathbf{r}'}^3}(\sin\Theta\cos\Theta),
\end{equation}
where $\mathbf{r'}$ is the position vector of the nucleus with the electron as origin and $\Theta$ is the angle that this vector forms with the magnetic field. The optimal resonance is given by $n=3$, which results in $\alpha\approx 0.72$. The shortest resonance ($n=1$) gives $\alpha\approx 0.37$. Under this dynamics, if we consider that the NV is initially polarized and both the electron and the nucleus are initially in a thermal state, the polarization of the nucleus $P'$ evolves after a sequence as
\begin{align}\label{eq:pol_transfer_NV_e_n}
    P'(\tau) &= \Tr\left[e^{-iH_{\text{eff}}\tau}\rho(0)e^{iH_{\text{eff}}\tau}2I_\text{z}\right] \nonumber\\    &=\frac{4 \mathcal{A}_\text{zz}^2 \left(\alpha \mathcal{B}_{\perp}\right)^2 }{\left[\mathcal{A}_\text{zz}^2 + \left(\alpha \mathcal{B}_{\perp}\right)^2\right]^2}\sin^4\left[\frac{\sqrt{\mathcal{A}_\text{zz}^2 + \left(\alpha \mathcal{B}_{\perp}\right)^2} }{8}\tau\right].
\end{align}
Whether the polarization is built on the nuclear $\ket{\uparrow}_n$ state or the $\ket{\downarrow}_n$ state (eigenstates of the $I_\text{z}$ operator) depends on the chosen harmonic. On the other hand, the polarization of the electron ($P_{\text{e}^-}$) also undergoes oscillations, at twice the frequency of the nuclear polarization, 

\begin{align}\label{eq:pol_transfer_NV_e}
    P_{\text{e}^-}(\tau) &= \Tr\left[e^{-iH_{\text{eff}}\tau}\rho(0)e^{iH_{\text{eff}}\tau}2E_\text{z}\right] \nonumber\\
    &=-\frac{\mathcal{A}_\text{zz}^2}{\mathcal{A}_\text{zz}^2 + \left(\alpha \mathcal{B}_{\perp}\right)^2}\sin^2\left[\frac{\sqrt{\mathcal{A}_\text{zz}^2 + \left(\alpha \mathcal{B}_{\perp}\right)^2} }{4}\tau\right].
\end{align}
Therefore, when the nucleus achieves it maximum polarization, the electron is back to being unpolarized. Equation~\eqref{eq:pol_transfer_NV_e_n} shows that the polarization transfer is maximized when the coupling between the NV and the electron equals the coupling between the electron and the nucleus (in absolute values) times the numerical constant $\alpha$. If we consider an electron right on top of the NV, at a distance $\abs{\mathbf{r}}$ such that $\theta=0$, the condition for the polarization transfer to be maximum is met for a nucleus located at a distance $\abs{\mathbf{r}'}$ from the electron
\begin{equation}
    \abs{\mathbf{r}'}\simeq \left(\frac{3\alpha\gamma_\text{n}}{4\gamma_\text{e}}\right)^{1/3}\abs{\mathbf{r}}
\end{equation}
where the $\Theta$-dependent term has been averaged over all possible orientations. For instance, if the pulse spacing is chosen such that $n=1$ in Eq.~\eqref{eq:tau} (which implies $\alpha\simeq 0.37$) and the polarization is transferred from an NV located 3.5 nm deep from the surface to a Hydrogen nucleus, the protocol is optimal if the nucleus is located at a distance $\abs{\mathbf{r}'}\simeq 2.6$ \r{A} from the dangling bond. In the case of an organic system with proton density $\rho_\text{N}\sim 50$ nm$^{-3}$ ~\cite{Tetienne2021}, there are in average two protons contained in the semi sphere of radius $\abs{\mathbf{r}'}=2.6$ \r{A}. Nevertheless, if this condition is not met and the polarization transfer is not maximized, the NV can be reinitialized and the sequence repeated until a higher level of polarization is achieved.

The use of the electron as a mediator presents an immediate advantage due to the electron's significantly larger gyromagnetic ratio compared to typical nuclei combined with the electron’s proximity to the diamond surface. As a result, the coupling constants $\mathcal{A}_\text{zz}$ and $\mathcal{B}_\perp$ are two to three orders of magnitude larger than those between the NV and the nucleus in a similar configuration (see Fig.~\ref{fig:sequence}(c)). Consequently, polarization buildup is significantly faster when utilizing the intermediate electron, compared to direct polarization transfer from the NV to the nucleus. Furthermore, in the direct PulsePol technique, surface electrons can introduce decoherence, whereas in our double-channel sequence, the surface electrons are controlled coherently. These advantages have important implications for improving the efficiency and sensitivity of experiments. 

\section{Results}\label{section:results}
\subsection{Discrete dynamical description}\label{section:III}

To illustrate our results, we begin by considering an idealized scenario where an NV center interacts with a single electron. In this case, if the system is perfectly coherent, polarization can be transferred between the electron and the NV following Eq.~\eqref{eq:pol1} at a rate only limited by the strength of their magnetic dipole coupling. Robustness is a crucial criterion for good polarization schemes~\cite{Korzeczek}, as the protocols need to be resistant to errors in the driving fields such as resonance offsets caused by the different strain conditions of each NV, leading to an effective change on $D$, and amplitude fluctuations of each delivered pulse. The former leads to detuning precessions ($\Delta_\text{NV}\neq 0$) in Hamiltonian~\eqref{eq:H_RWA} that accumulate during the free evolution, while the latter produces errors in the Rabi frequency  ($\Tilde{\Omega} = \Omega+\Omega_\text{error}$), where $\Omega$ represents the Rabi frequency generating ideal pulses for a given pulse duration. These errors generate rotations that differ from perfect $\pi$ or $\pi/2$ pulses.
%
%%%%%%%%%%%%%%%%%%%%%%%%%%%%%%%%%%%%%%%%%%
\begin{figure*}[t]
    \centering
    \includegraphics[width=0.9\linewidth]{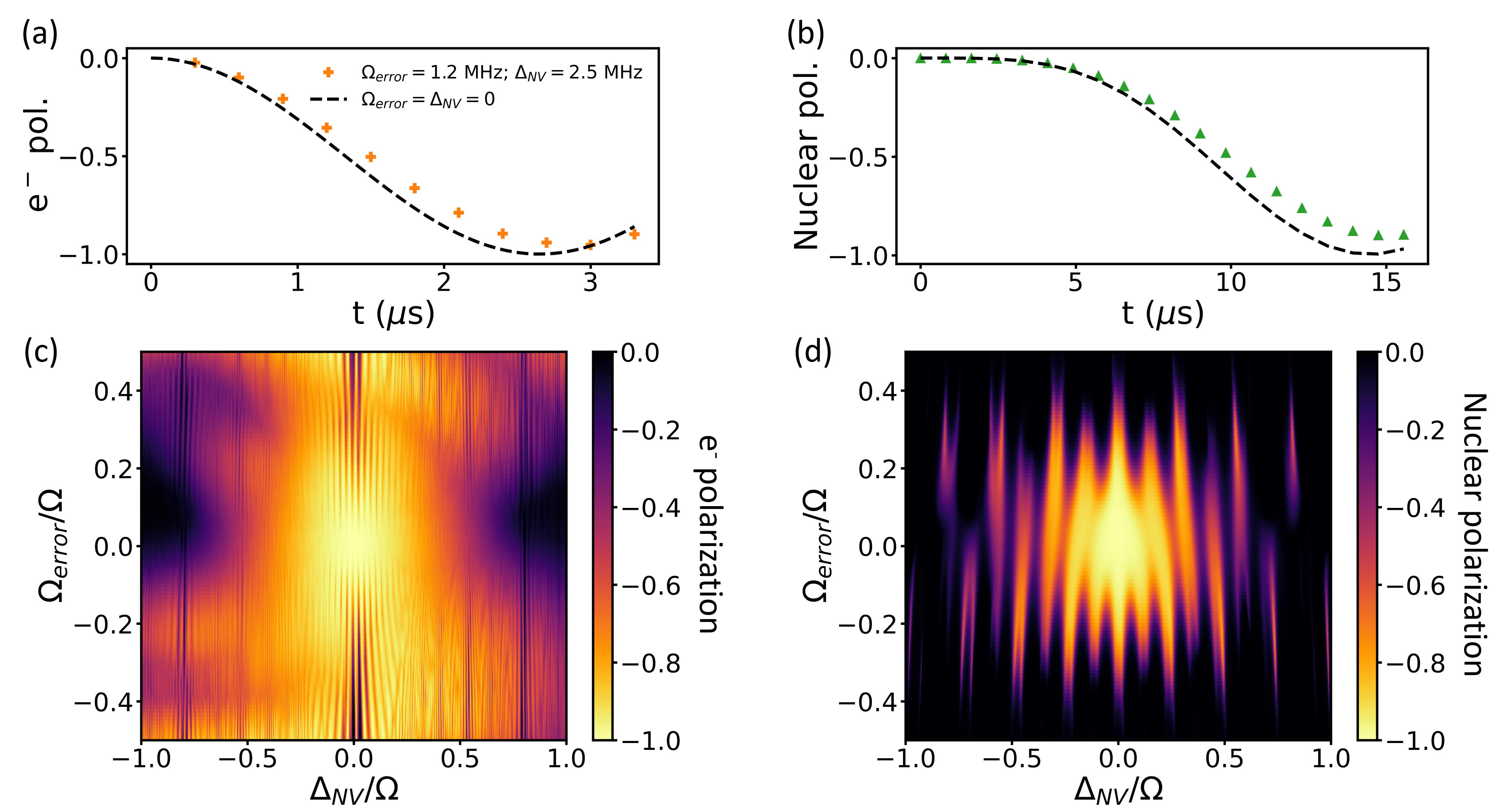}
  \caption{Robustness simulations of the double-channel PulsePol. (a) Polarization transfer to a single electron spin under perfect conditions (black dashed line) and under pulse errors (orange crosses) that include a detuning error $\Delta_\text{NV}=(2\pi)$2.5 MHz in the driving field of the nitrogen-vacancy (NV) spin and Rabi frequency errors $\Omega_\text{error}=(2\pi)$1.2 MHz in both driving fields. The NV-electron coupling is $\mathcal{A}_\text{zz}=(2\pi)$0.4 MHz and the Rabi frequency $\Omega=(2\pi)$20 MHz. (b) Polarization transfer to a $^1$H nucleus using an electron as mediator under perfect conditions (black dashed line) and under the same pulse errors and parameters as in (a) (green triangles). The NV-electron and the electron-nuclear couplings are $\mathcal{A}_\text{zz}=\alpha\mathcal{B}_\perp=(2\pi)0.1$MHz. (c-d) For the same parameters as (a-b), maximum polarization transfer versus $\Delta_\text{NV}$ and $\Omega_\text{error}$. The color bars represent the polarization acquired by the target particle, quantified by $2\langle E_z\rangle$ and $2\langle I_z\rangle$, respectively, being $-1$ the maximum attainable polarization. }
  \label{fig:robustness}
\end{figure*}
%%%%%%%%%%%%%%%%%%%%%%%%%%%%%%%%%%%%%%%%%%
%

Figure \ref{fig:robustness}(a) presents numerical simulations of the polarization, quantified by $\langle 2E_\text{z} \rangle$, acquired by an electron under the proposed scheme near an initially fully polarized NV spin in the presence of errors, which is then compared with the faultless transfer (dashed line), which is described by Eq.~\eqref{eq:pol1}. The Rabi frequencies are taken such that $\Omega_1=\Omega_2\equiv\Omega$, and thus its error affects both channels. The selected parameters (indicated in the caption) are characteristic of a near-surface NV center spin and a somewhat imprecise pulse scheme. As anticipated from the PulsePol structure, the errors have a limited impact on polarization transfer. Even in the presence of these errors, the electron still attains 95\% of the total polarization, albeit with a slightly extended duration compared to the error-free sequence (approximately 14\% slower). In Fig.~\ref{fig:robustness}(b), we demonstrate the robustness of our protocol by showing the resulting polarization transfer for different values of $\Delta_{\text{NV}}$ and $\Omega_\text{error}$. Similar to the PulsePol applied to a single channel, the acceptable Rabi amplitude error and resonance offset scale with the Rabi frequency (see Supplementary Note 4). We observe that efficient polarization transfer can still be achieved for a wide range of errors. However, when simulating the polarization of a single nuclear spin using the proposed protocol, as illustrated in Figs. \ref{fig:robustness}(c) and \ref{fig:robustness}(d), we observe that the mediation of the interaction through the electron introduces complexities. This additional element in the polarization transfer process makes it more sensitive to errors, wherein smaller inaccuracies in the driving strength or frequency can hinder the effectiveness of the sequence. As shown in Fig.~\ref{fig:robustness}(c), when subjected to the same faulty parameters as in the previous case, the sequence maintains about 90\% of its effectiveness with only a 6\% decrease in speed. The results in Fig.~\ref{fig:robustness}(d) shows that, despite the heightened sensitivity, the sequence remains viable for a broad range of detunings and even in the presence of faulty pulses, demonstrating the robustness of the double-channel PulsePol sequence for direct polarization transfer to a target spin and highlighting the importance of careful control of experimental parameters when aiming to polarize additional nuclear spins using the first target spin as the mediator.

To further study the case of the electron being used as mediator to transfer polarization outside of the diamond lattice, we consider an environment containing only one electron and multiple target spins, since the typical nuclear densities in the target region is much larger than the density of dangling bonds at the surface. In a system consisting of an NV, an electron and multiple target spins, polarization is still transferred following Eq.~\eqref{eq:pol_transfer_NV_e_n}, with the  modification
\begin{equation}\label{eq:A+B}
\mathcal{A}_\text{zz}^2 + \left(\alpha \mathcal{B}_{\perp}\right)^2\mapsto \mathcal{A}_\text{zz}^2 + \left(\alpha \mathcal{B}_0\right)^2,
\end{equation}
where
\begin{equation}
\mathcal{B}_0^2=\sum_i^\text{nuclei}\left(\mathcal{B}_\perp^{(i)}\right)^2.
\end{equation}
Since at most a single spin state provided by the NV is shared among an ensemble of nuclear spin, in order to build up polarization, the NV must be frequently reinitialized, and the protocol must be repeated multiple times. So as to maximize polarization transfer rate, the sequence must be repeated $N$ times before reinitializing the NV, until each nucleus acquires it maximum possible polarization. It is important to note that a continuous optical driving of the NV spin into its $\ket{0}$ state would result in a quantum Zeno-like effect that slows down the effective polarization rate. After $N$ cycles before reinitializing, with duration $T=2N\tau$, where $\tau$ is the duration of a single sequence, each nucleus acquires at most
\begin{equation}\label{eq:P(t+T)}
    P^{(i)}(t+T) = P^{(i)}(t) + \left[1-P^{(i)}(t)\right]T u_i(T),
\end{equation}
where $u_i(T)$ is the cooling rate of nucleus $i$, defined according to Eq.~\eqref{eq:pol_transfer_NV_e_n} as
\begin{equation}
    u_i(T)=\frac{4 \mathcal{A}_\text{zz}^2 \left(\alpha \mathcal{B}_{\perp}^{(i)}\right)^2 }{T\left[\mathcal{A}_\text{zz}^2 + \left(\alpha \mathcal{B}_0\right)^2\right]^2} \sin^4\left[\frac{\sqrt{\mathcal{A}_\text{zz}^2 + \left(\alpha \mathcal{B}_0\right)^2} }{8}T\right].    
\end{equation} 
Treating Eq.~\eqref{eq:P(t+T)} as a differential equation, for $T$ sufficiently small compared to the cooling rate, the polarization buildup can be approximated by an exponential,
\begin{equation}\label{eq:P_exp}
    P^{(i)}(t)\approx P^{(i)}(0)+\left[1-P^{(i)}(0)\right]\left[1-e^{-u_i(T)t}\right].
\end{equation}
%
%
%%%%%%%%%%%%%%%%%%%%%%%%%%%%%%%%%%%%%%%%%%
\begin{figure}[t!]
    \centering
    \includegraphics[width=0.95\linewidth]{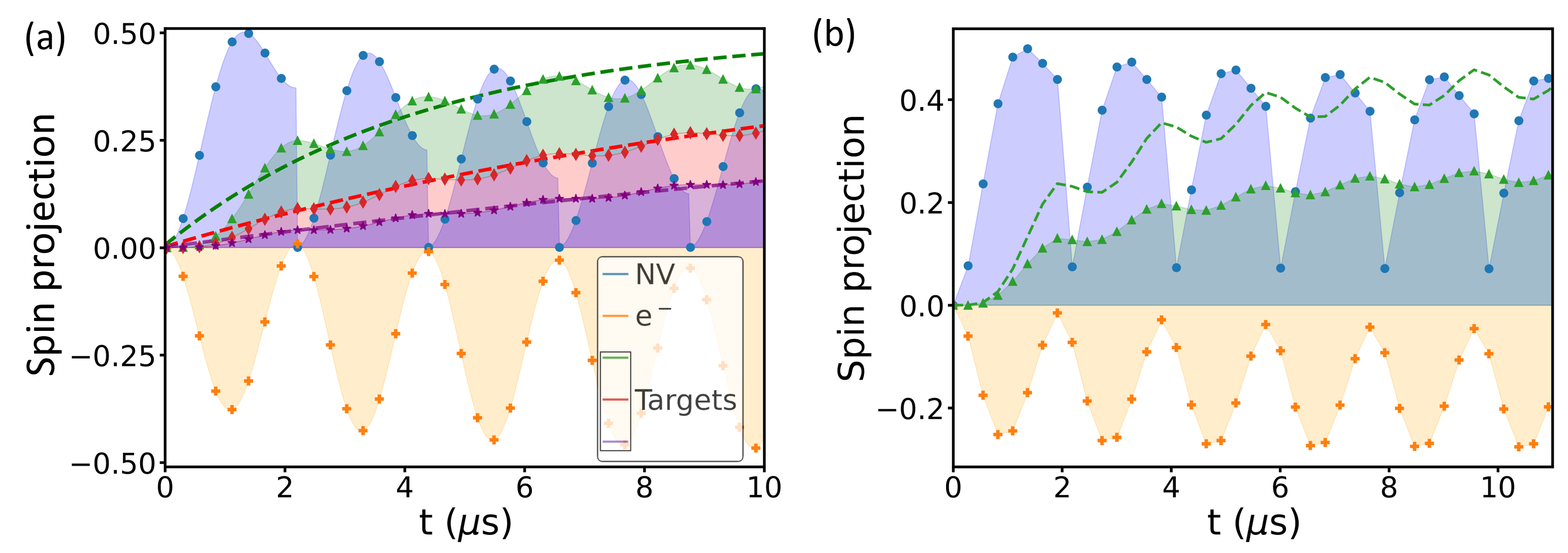}
   \caption{Theoretical overview of nitrogen-vacancy (NV)-electron-nuclei polarization transfer mechanism for the case in which all elements are treated as discrete entities. Symbols represent polarization of each element at the end of a double-channel PulsePol sequence. (a) Dynamical simulation of one NV, an electron and 3 target $^1$H nuclei. Dashed lines denote the analytical expression given by Eq.~\eqref{eq:P_exp}. The NV and electron interact with a coupling $\mathcal{A}_\text{zz}=(2\pi)0.9$ MHz, while the nuclei interact with the electron with couplings $\mathcal{B}_\perp=$ $(2\pi)1$ MHz, $(2\pi)0.6$ MHz and $(2\pi)0.4$ MHz. The magnetic field is set to 430 G and the pulses are considered to be ideal and instantaneous. The interpulse spacing corresponds to $n=3$, and the NV is re-initialized every 8 sequences. (b) Dynamical simulation of one NV, an electron and a target $^1$H nucleus (first nucleus of the previous simulation) with decoherence. The relaxation and coherence times are $T_1^\text{NV}=1$ ms, $T_2^\text{NV}=10$ $\mu$s, $T_1^{e^-}=30$ $\mu$s, $T_2^{e^-}=1$ $\mu$s, $T_1^\text{n}=1$ s, $T_2^\text{n}=1$ ms. In this simulation, we assume that the NV is 80\% polarized in the $\ket{0}$ state. The green dashed line represents the polarization acquired in a fully decoherence-free environment. The inter-pulse spacing corresponds to $n=3$, and the NV is re-initialized every 7 sequences.} 
  \label{fig:model}
\end{figure}
%%%%%%%%%%%%%%%%%%%%%%%%%%%%%%%%%%%%%%%%%%
%
The results of a simulated realisation of the sequence in a system composed by an NV, an electron and 3 nuclei are depicted in Fig.~\ref{fig:model}(a). The y-axis of the figure represents the mean z-component of the spin for each of the particles ($\langle S_z\rangle$, $\langle E_z\rangle$ and $\langle I_z^{(i)}\rangle$ for the NV, the electron, and the nuclei, respectively). In the polarization transfer process, the three target nuclei exhibit distinct polarization rates that depend on their individual dipole-dipole couplings to the electron. The mean dynamics of each nucleus are effectively described by Eq.~\eqref{eq:P_exp}, represented by the corresponding dashed lines in the figure. Once the nuclei reach a maximum in polarization, the NV center is re-initialized via optical pumping to the $\ket{0}$ state. The abrupt declines in NV dynamics at specific time points (approximately 2 µs, 4 µs, 6 µs, etc.) are a consequence of this optical pumping event.

The previous description does not take into account the relaxation rates of the target nuclei, the nuclei-nuclei interaction that can give rise to polarization diffusion through the sample, or decoherence rates of the NV or electron. In fact, one of the biggest obstacles of this protocol is the extremely short coherence time of the electron, in the order of 1 $\mu$s. To include the effect of decoherence, in Fig.~\ref{fig:model}(b) we numerically simulate the dynamics of an NV center interacting with an electron and a nuclear spin under typical relaxation and dephasing rates using a quantum master equation that incorporates these rates (see Supplementary Note 5). As shown in green dashed line, the polarization acquired under ideal decoherence-free conditions is significantly higher than the polarization obtained under typical decoherence rates. However, even with the presence of decoherence, the double-channel PulsePol sequence demonstrates significantly enhanced polarization transfer compared to the direct application of the single-channel PulsePol sequence to a nucleus positioned according to the selected couplings. Notably, the direct method, characterized by a coupling between the NV and the nucleus in the range of KHz, leads to polarization transfer times on the order of milliseconds. Consequently, the acquired polarization by the nucleus is deemed negligible within the simulation's chosen time scale, which is on the order of some microseconds.

\subsection{Continuum description}\label{section:IV}Accurately modeling these systems typically requires considering regions of tens or hundreds of nanometers in size, which contain around 10$^{10}$ proton spins and 10$^{20}$ couplings in a typical organic target sample. This makes discrete dynamical modeling impractical. However, the relatively high target densities (approximately 50 nm$^{-3}$) compared to the NV-target standoffs (approximately 5 nm) allow for a continuum description of the system. If a single spin interacts with an ensemble of independent spins with density $\rho_\text{N}(\mathbf{R})$, it will still exchange polarization coherently with the ensemble as it was described in Eq.~\eqref{eq:A+B}, but now $\mathcal{B}_0$ is given by~\cite{Broadway2018}
\begin{equation}\label{eq:A0}
\mathcal{B}_0^2 = \int\rho_\text{N}(\mathbf{R})\mathcal{B}_\perp(\mathbf{R})^2d^3\mathbf{R},
\end{equation}
where $\mathcal{B}_\perp(\mathbf{R})$ is the coupling strength given by Eq.~\eqref{eq:Bzx} for a spin located at position $\mathbf{R}$, and the integral is taken over the target region external to the diamond surface. 

In the approach taken in this section, we focus on the polarization of an infinitesimal fixed volume located at position $\mathbf{R}$, with individual spins potentially moving in or out, rather than tracking the trajectories and polarization of individual particles. According to Broadway et al.\cite{Broadway2018}, the system is modelled through a convection-diffusion equation with spatially dependent cooling rate $u(\mathbf{R})$,
\begin{align}\label{eq:P}
\frac{\partial}{\partial t} &P(\mathbf{R},t) = \nonumber\\
&u(\mathbf{R})\left[1-P(\mathbf{R},t)\right] - \Gamma_{1,\text{n}}P(\mathbf{R},t)+D\nabla^2P(\mathbf{R},t).
\end{align}
where $\Gamma_{1,n}$ is the spin-lattice relaxation of the nuclear spins and $D = D_{dp}+D_{sp}$ is the total effective diffusion constant due to processes associated with nuclear dipole-dipole (dp) mediated, and spatial (sp) diffusion respectively. In a liquid sample, this diffusion term can capture molecular diffusion if it is slow enough for the flip-flop dynamics described in the previous section to remain approximately valid. In a solid sample, it can describe dipole-mediated spin diffusion. The cooling rate describes the polarization buildup after a cycle consisting of initializing the NV spin to $\ket{0}$ and applying the protocol for a duration $T=2N\tau$. Therefore, the cooling rate depends on the specific protocol used. For direct polarization transfer, the cross-relaxation (CR)~\cite{Broadway2018} and NOVEL methods have been shown to achieve larger cooling rates than the PulsePol method, thanks to their relatively larger flip-flop coupling strength~\cite{Hall2020}. However, PulsePol is less sensitive to NV spin dephasing, which allows it to operate in the coherent regime for external samples. Furthermore, as it will be shown next, by applying the double-channel PulsePol sequence, the cooling rate can be significantly increased.

We consider a semi-infinite slab of H$^1$ spins in frozen water with uniform density ($\rho_\text{N}(\mathbf{R})=66$ nm$^{-3}$) and placed on a flat diamond surface. An NV center is located at the origin, beneath the surface. A surface electron, which is used as a polarization mediator, is placed at position $\mathbf{r}$. In this scenario, the cooling rate is given by the same expression as it is used in Eq.~\eqref{eq:P(t+T)}, substituting the coupling with the $i$-th spin $B_\perp^{(i)}$ by a position-dependent coupling $B_\perp(\mathbf{R})$. The expression for the cooling rate can then be slightly modified to incorporate some practical considerations affecting the protocol implementation. For instance:
\begin{enumerate}[label=(\roman*)]
\item The surface electron and the NV have a finite coherence time that severely limits polarization transfer. In particular, the electron has a dephasing time of approximately $1$ $\mu$s, while the NV's dephasing time is approximately $10$ $\mu$s, as reported in Refs.~\cite{Lukin2014,Healey2021}. The PulsePol protocol is specifically designed to be resistant to quasistatic dephasing, which suggests that the results obtained in our simulations may represent a conservative estimate of its performance. Nevertheless, the efficacy of polarization transfer is still affected by faster fluctuations in the system, potentially leading to reduced transfer efficiency. As a rough approximation, these coherences can be taken into account by including a damping factor $e^{-(\Gamma_2^{e^-}+\Gamma_2^{\text{NV}})T}$ in the expression of $u(\mathbf{R})$\cite{Healey2021}, where $\Gamma_2^{e^-}$ and $\Gamma_2^{\text{NV}}$ are the dephasing rates (inverse of the dephasing time) of the electron and the NV, respectively.
\item In our analysis, we made the idealized assumption that the spin of the NV center could be perfectly initialized into the pure spin $\ket{0}$ state. However, in practice, the initialization fidelity of the NV center is finite, with a value denoted by $\mathcal{F}_\text{NV} <1$. For single NV centers in bulk diamond, the typical observed value of the initialization fidelity is around $\mathcal{F}_\text{NV} \approx 0.8$ for short duration pulses~\cite{Robledo2011}. This finite fidelity has a direct impact on the degree of polarization transferred to the electron, which in turn can diminish the efficiency of the hyperpolarization process. To account for the effect of $\mathcal{F}_{NV}$, we straightforwardly incorporate it into the cooling rate $u(\mathbf{R})$ as a multiplicative factor. 
\item In addition to the finite initialization fidelity, the process is not instantaneous but instead requires a finite amount of time. Specifically, the initialization typically involves at least $1$ $\mu$s of optical pumping under high laser intensity, followed by $1$ $\mu$s for the NV center to relax to its ground state~\cite{Manson2006}, before the polarization transfer protocol can be applied. To incorporate this delay into our model, we can introduce a dead time $t_d = 2$ $\mu$s, which can be added to the time $T$ in the denominator of the expression for the cooling rate.
\end{enumerate}
Combining the above factors, the cooling rate after applying the protocol for a single cycle of duration $T$ can be approximated as
\begin{equation}
    u(\mathbf{R}) = \frac{\mathcal{F}_\text{NV}e^{-(\Gamma_2^{e^-}+\Gamma_2^{\text{NV}})T}}{T+t_d}v(\mathbf{R}),
\end{equation}
where $v(\mathbf{R})$ is calculated as
\begin{align}
    v&(\mathbf{R})=\nonumber\\
    &\frac{2\mathcal{A}_\text{zz}(\mathbf{r})^2\left[\alpha\mathcal{B}_{\perp}(\mathbf{R})\right]^2}{\left[\mathcal{A}_\text{zz}(\mathbf{r})^2+\left(\alpha\mathcal{B}_{0}\right)^2\right]^2}\sin^4\left[\frac{\sqrt{\mathcal{A}_\text{zz}(\mathbf{r})^2 + \left(\alpha \mathcal{B}_{0}\right)^2} }{8}T\right],
\end{align}
and $\mathcal{B}_0$ is calculated using Eq.~\eqref{eq:A0} with the $\mathcal{B}_{\perp}$ coupling term from Eq.~\eqref{eq:Bzx}.
\begin{figure*}[t]
    \centering
    \includegraphics[width=1\linewidth]{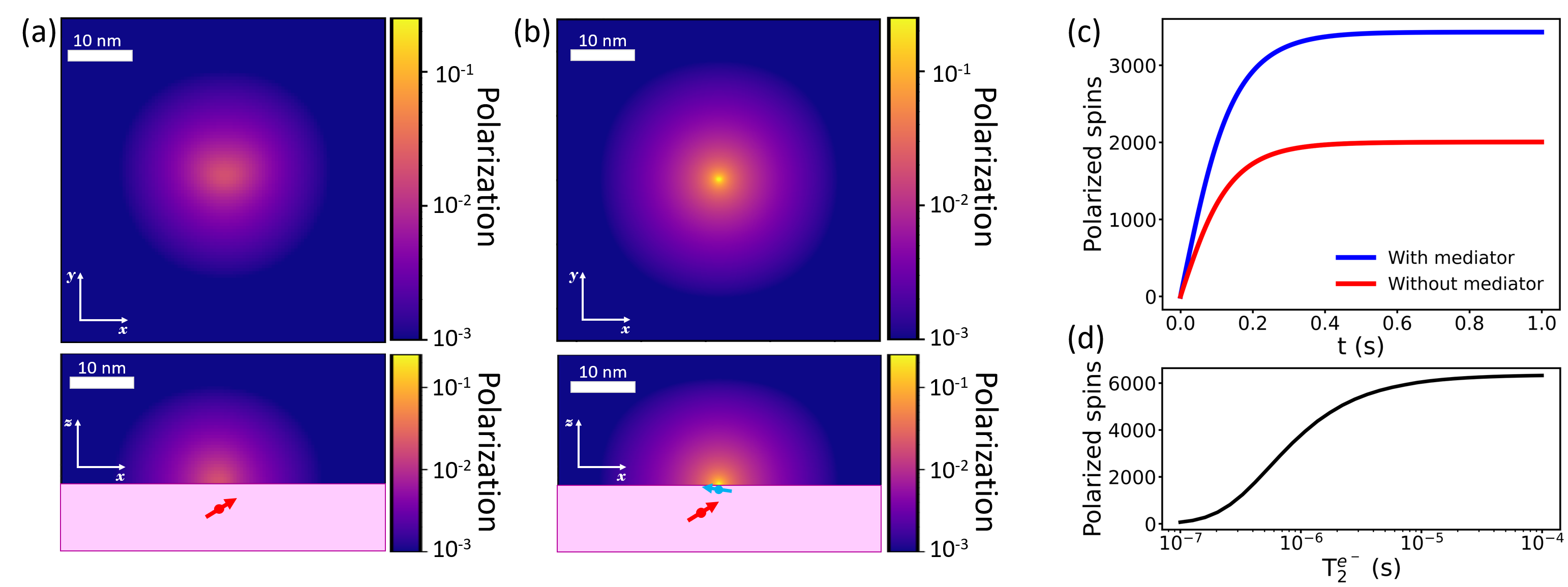}
  \caption{Numerical simulation of the hyperpolarization process. The nitrogen-vacancy (NV) spin has a depth $d_\text{NV} = 3.5$ nm and is oriented at an angle $\theta_\text{NV} = 54.7^\text{o}$ in the xz plane, which is the most commonly found angle as it corresponds to a (100)-oriented diamond surface. The nuclear spins ($^1$H) have a density $\rho_\text{N} = 66$ nm$^{-3}$ and a relaxation rate $\Gamma_{1,n} = 1$ s$^{-1}$. (a) Steady-state polarization maps in the xy plane at the diamond-sample interface and (b) in the xz plane including the NV center in the direct hyperpolarization case using the PulsePol. (c-d) Steady-state polarization maps when the electron is used as polarization mediator. (e) Number of polarized spins, given by Eq.~\eqref{eq:N} in the conditions of figures (a-d). (f) Calculated number of polarized spins as a function of the coherence time of the electron $T_2^{e^-}$. }
  \label{fig:XYZ_NV_e}
\end{figure*}
%%%%%%%%%%%%%%%%%%%%%%%%%%%%%%%%%%%%%%%%%%
%

Our analysis considers an NV center located at a depth of $d = 3.5$ nm beneath the diamond surface, with the quantization axis oriented at an angle of $54.7^\text{o}$ with respect to the surface's normal. This orientation is chosen strategically to minimize the average coupling strength among surface electrons that might potentially interact. The specific angle of $54.7^\text{o}$ ensures the lowest possible average coupling strength among these surface electrons.
We introduce a single mediator electron positioned on the diamond surface. Notably, the coupling between the NV center and the electron vanishes at $\theta=54.7^o$ (see Eq.\eqref{eq:Azz}). Hence, we deliberately place the electron at a position where it is not directly above the NV center. Through our analysis, we find that the optimal duration for maximizing the polarization rate is approximately $T_0 \simeq 0.6$ $\mu$s. Under a static magnetic field intensity of $B = 390$ G, this corresponds to a single PulsePol sequence block with an interpulse spacing corresponding to $n = 1$ in Eq.~\eqref{eq:tau}.

We contrast the results obtained through the use of the electron mediator with those arising from direct polarization transfer from the NV center to the external nuclei utilizing the PulsePol sequence. In this setting, and accounting for all of the practical constraints that were mentioned earlier, the cooling rate can be expressed as 
\begin{equation}
    u^{\text{PP}}(\mathbf{R}) = \frac{\mathcal{F}_{\text{NV}}e^{-\Gamma_2^{\text{NV}}T'}}{T'+t_d}\frac{\left[\mathcal{B}'_{\perp}(\mathbf{R})\right]^2}{\left(\mathcal{B}_{0}'\right)^2}\sin^2\left[\frac{\alpha'\mathcal{B}'_{0}}{4}T'\right].
\end{equation}
Notice that in the direct polarization transfer scheme, the constant $\alpha'$ may differ from the $\alpha$ value used in the double-channel sequence, and the coupling $\mathcal{B}'_0$ is computed based on the position vector of the nuclei relative to the NV center. Consequently, $\mathcal{B}'_0$ is much smaller than $\mathcal{B}_0$, as the target sample is located much farther away from the NV center compared to the mediator electron. Under the same conditions as the previous example, we find that the optimal duration for maximizing the polarization rate in this scenario is approximately $T_0' \simeq 7$ $\mu$s. This corresponds to a duration equivalent to approximately 4 PulsePol sequence blocks with interpulse spacing corresponding to $n=3$ in Eq.~\eqref{eq:tau}, assuming the same magnetic field intensity as before. 

To simulate the hyperpolarization process described by Eq.~\eqref{eq:P}, we assume that the target nuclei have a relaxation constant $\Gamma_{1,\text{n}}=1$ s$^{-1}$. The diffusion constant is computed using the approximate expression for a cubic lattice, $D \simeq 0.22\mu_0\hbar\gamma_\text{n}^2\rho_\text{N}^{1/3}/4\pi$ \cite{Cheung1981}, resulting in $D\simeq 670$ nm$^2s^{-1}$. The presence of diffusion causes the polarization to spread outwards from the source, ultimately reaching a steady-state configuration in a time of approximately the relaxation time $T_{1,\text{n}} = 1/\Gamma_{1,\text{n}}$, or even faster if the cooling rate $u(R)$ is much larger than the relaxation rate $\Gamma_{1,\text{n}}$. The spatial extent of the polarization can be estimated to be on the order of $\sqrt{D_\text{n} T_{1,\text{n}}}\approx 30$ nm. Polarization maps simulated for the described situations are plotted in Figs.~\ref{fig:XYZ_NV_e}(a) and (b) for the direct polarization transfer from NV to nuclei and Figs.~\ref{fig:XYZ_NV_e}(c) and (d) for the electron-mediated polarization transfer case. The use of the electron as mediator leads to a larger total polarization compared to direct transfer to the external nuclei using PulsePol. In particular, the polarization is concentrated more strongly near the mediator electron, as its sphere of influence is much smaller due to its proximity to the surface. This localized polarization enhancement could be advantageous for hyperpolarization in highly diffusive samples, where efficient transfer of polarization to nearby nuclei is crucial before it diffuses away. To quantify the efficiency of the polarization process, a useful metric is the effective number of polarized spins, which is defined as
\begin{equation}\label{eq:N}
    N(t)=\rho_\text{N}\int P(\mathbf{R},t)d^3\mathbf{R}.
\end{equation}
Under the conditions described in Fig.~\ref{fig:XYZ_NV_e}, we observe that the steady-state population of polarized nuclei is approximately 2000 without a mediator, while it increases to $N \approx 3100$ when using the electron-mediated polarization transfer protocol. This improvement is due to the higher cooling rate provided by the electron-mediated protocol. While our initial analysis anticipates significantly accelerated polarization transfer rates when utilizing the intermediate electron, the observed enhancement factor in the final results is only approximately 1.5 compared to the direct protocol. This outcome arises from the interplay of two primary factors: decoherence and spatial constraints. Decoherence poses a substantial influence on the efficiency of the polarization transfer process. Notably, the electron experiences faster decoherence rates compared to the NV, impacting the maintenance and coherence of the polarized state. This difference in decoherence dynamics contributes to the modest enhancement observed in the final results. Furthermore, the spatial constraints imposed by the proximity of the electron to the diamond surface play a pivotal role. The radius of the sphere of action of the polarizing agent, delineating the region within which it exhibits significantly larger couplings with nearby nuclei, is approximately equal to its depth inside the diamond crystal. Given the electron's close proximity to the surface, its sphere of action is comparatively smaller than the NV's sphere of action. Consequently, nuclei in immediate proximity to the electron dominate the interaction, resulting in a localized polarization effect that limits the transfer to other nuclei. 

Fig~\ref{fig:XYZ_NV_e}(e) shows the dynamic polarization buildup for both the direct and electron-mediated protocols. In the electron-mediated case, the number of polarized spins initially grows faster due to the electron's higher cooling rate. However, the high local polarization around the electron acts as a blocking mechanism, limiting the polarization growth. Nonetheless, diffusion allows for polarization at distances beyond that reachable via the dipole-dipole interaction, enabling a higher total polarization to be reached, and resulting in a larger number of polarized spins in the steady-state. In both cases, the steady-state is attained at approximately the same time.

As mentioned, decoherence is the most significant factor affecting the protocol. Figure~\ref{fig:XYZ_NV_e}(f) illustrates the effect of the electron's decoherence. Increasing the coherence time of the electron to around $4$ $\mu$s can result in a twofold increase in the number of polarized spins. Although this value of the coherence time is currently beyond the reported value by Sushkov et al.\cite{Lukin2014}, optimizing diamond surface preparation and utilizing the intrinsic decoupling scheme of the PulsePol could result in a significant coherence boost. Beyond this point, further enhancements in the electron's coherence may not produce remarkable improvements as the NV's decoherence starts to limit the effectiveness of the protocol.

\section{Conclusion}\label{section:V}
We introduced a double-channel sequence with which, by applying the PulsePol sequence to the NV and the target spin separately, it is possible to transfer polarization to particles with spin regardless of their Larmor frequency. Importantly, while we have predominantly discussed the role of electrons as mediators in this work, it is crucial to highlight the broad applicability of the proposed sequence. Our method is not exclusive to electrons; rather, it extends to any spin, even spins with a high Larmor frequency. This versatility opens new possibilities for hyperpolarization techniques, especially for spins that pose challenges for conventional sequences. The main advantages of this double-channel protocol are that the restriction on the inter-pulse separation to fulfill the resonance with the target Larmor frequency is lifted, and that the full coupling strength contributes to the polarization rate. The sequence is robust to errors in the driving fields such as detunings or pulse errors. All this makes the double-channel PulsePol sequence a promising candidate for a variety of applications, including polarizing electrons in moderate magnetic fields or nuclear spins in intense magnetic fields. Moreover, this procedure could be extended to the near-equivalent regime of parahydrogen-induced polarization, where PulsePol has already been shown to be effective~\cite{Korzeczek}. Future work can explore the use of this sequence applied to parahydrogen polarizing agents to polarize nuclei at high fields, a method that has already been demonstrated in recent studies at low fields through signal amplification by reversible exchange~\cite{Arunkumar2021}.

Additionally, this sequence has the ability to transfer polarization to nearby nuclear spins using the electron as a mediator by meeting the PulsePol resonance condition on the inter-pulse spacing. While our initial analysis suggested the potential for substantially improved polarization transfer capabilities using this approach, leveraging the anticipated strengthened couplings between the NV center, the electron mediator, and the target nuclei, the realized enhancement factor in the final results demonstrates a more modest increase, reaching approximately 1.5 compared to the direct protocol. This nuanced outcome is intricately linked to the dynamic interplay of decoherence mechanisms and spatial constraints. Despite these challenges, our approach stands out as superior to other state-of-the-art methods, showcasing the potential of the double-channel PulsePol sequence. Moreover, in a more realistic scenario, the presence of multiple dangling bonds interacting with the NV could serve as additional polarization transfer mediators. Considering the polarization's localized nature in close proximity to the electrons, the inclusion of multiple electrons might offer further benefits to the protocol. However, a more in-depth investigation is warranted to ascertain the validity of this assumption.

\section*{Acknowledgements}
This version of the article has been accepted for publication, after peer review,
but is not the Version of Record and does not reflect post-acceptance improvements, or any
corrections. The Version of Record is available online at: https://doi.org/10.1038/s42005-024-01536-6. We acknowledge financial support from Spanish Government via the projects PID2021-126694NA-C22 and PID2021-126694NB-C21 (MCIU/AEI/FEDER, EU), the ELKARTEK project Dispositivos en Tecnolog\'i{a}s Cu\'{a}nticas (KK-2022/00062), the Basque Government grant IT1470-22, and by Comunidad de
Madrid-EPUC3M14. H.E. acknowledges the Spanish Ministry of Science,
Innovation and Universities for funding through the FPU
program (FPU20/03409). C.M.-J. acknowledges the predoctoral MICINN grant PRE2019-088519. J. C and E.T. acknowledge the Ram\'on y Cajal (RYC2018-025197-I and RYC2020-030060-I) research fellowship.

\section*{Data availability}
All data generated or analysed during the simulations of study are included in this published article and its supplementary information files.

\section*{Code availability}
The code employed to generate the simulations presented in this work is available upon request. Interested parties may inquire about access to the code by contacting the corresponding author.

\section*{Author contributions}
H.E. performed the analytical calculations. H.E., I.P., and C.M.-J. carried out the simulations. C.M.-J. and J.C. designed the microwave sequence. The paper was written by H.E., C.M.-J., J.C., and E.T., with valuable input and discussions from R.P., and P.A. J.C. and E.T. envisioned and coordinated the project.

\section*{Competing interests}
The authors declare no competing interests.

\appendix

\bibliography{mainbib.bib}{}
\bibliographystyle{naturemag}
\end{document}